\documentclass{aastex701}

\usepackage[]{mdframed}
\usepackage{graphicx}
\usepackage{wrapfig}
\usepackage{enumitem}
\usepackage{natbib}

\usepackage{hyperref}
\hypersetup{
    colorlinks=true,
    citecolor=blue,
    linkcolor=red,
    filecolor=magenta,      
    urlcolor=cyan,
    }

\begin{document}

\title{The critical role of HST UV spectroscopy in constraining red supergiant
binary systems  \newline \footnotesize Building a Roadmap for Hubble science into the 2030s}

\author{Daniel Jadlovský}
\affiliation{Department of Theoretical Physics and Astrophysics, Faculty of Science, Masaryk University, Kotl\'a\v rsk\'a 2, 61137, Brno, Czech Republic}
\affiliation{European Southern Observatory (ESO), Karl-Schwarzschild Str. 2, 85748, Garching bei München, Germany}
\email[show]{jadlovsky@mail.muni.cz}  


\author{Andreas Sander}
\affiliation{Zentrum f\"ur Astronomie der Universit\"at Heidelberg, Astronomisches Rechen-Institut, M\"onchhofstr.\ 12-14, 69120 Heidelberg, Germany}
\affiliation{Interdisziplin{\"a}res Zentrum f{\"u}r Wissenschaftliches Rechnen, Uni. Heidelberg, Im Neuenheimer Feld 225, 69120 Heidelberg, Germany} 
\email{andreas.sander@uni-heidelberg.de}

\author{Lee Patrick}
\affiliation{Centro de Astrobiología (CSIC-INTA), Ctra. Torrejón a Ajalvir km 4, 28850 Torrejón de Ardoz, Spain}
\email{fakeemail4@google.com}

\author{Jiří Krtička}
\affiliation{Department of Theoretical Physics and Astrophysics, Faculty of Science, Masaryk University, Kotl\'a\v rsk\'a 2, 61137, Brno, Czech Republic}
\email{fakeemail6@google.com}

\author{Gemma González-Torà}
\affiliation{Zentrum f\"ur Astronomie der Universit\"at Heidelberg, Astronomisches Rechen-Institut, M\"onchhofstr.\ 12-14, 69120 Heidelberg, Germany}
\email{fakeemail5@google.com}

\author{Matheus Bernini-Peron}
\affiliation{Zentrum f\"ur Astronomie der Universit\"at Heidelberg, Astronomisches Rechen-Institut, M\"onchhofstr.\ 12-14, 69120 Heidelberg, Germany}
\email{fakeemail7@google.com}



\begin{abstract}
\noindent%
The Hubble Space Telescope (HST) has advanced our knowledge of stellar evolution thanks to its unique high-spectral resolution and high-sensitivity capabilities in the UV region, especially for hot massive stars. Within the field of red supergiants (RSGs), it has led to several major insights as well, such as, for example, characterizing the embedded hot companion in the VV Cephei system, the detection of surface mass ejection from Betelgeuse, and unveiling the rich population of RSG+B binaries in the Magellanic Clouds. We demonstrate that nowadays (and in the decade to come) we need HST more than ever, as for the first time, thanks to the significant advances of high-angular-resolution interferometric and astrometric techniques, we can resolve the orbital motion of stellar components in RSG systems. However, the long wavelengths are dominated by the cool RSGs. Thus, observing such companions in the UV with HST is the only way to constrain the properties and velocity of the hot companions, adding the required missing piece of information for constraining the orbital solution and evolutionary status of RSG systems. Such observations are critical for understanding the interaction and mass transfer in massive binary systems and their connection to the observed population of supernovae.
\end{abstract}



\section*{Introduction and Scientific Context}

\noindent
As strong sources of stellar feedback and chemical enrichment, massive stars are major drivers of cosmic evolution. At the end of their lives, massive stars collapse into neutron stars or black holes, often accompanied by a powerful supernova (SN) explosion. With their strong winds and as significant sources of chemical enrichment and dust production, \textbf{red supergiants} (RSGs) mark a particularly important evolutionary stage among massive stars and are key players in the cosmic matter cycle \citep[e.g.,][]{Bergemann+2012,Levesque2017}. 
The assumed properties and fate of RSGs in evolution and population synthesis modeling have wide-ranging impacts, affecting, e.g., the predicted dust budget, conditions for planet formation, and predicted SN statistics. The textbook picture of massive star evolution presents RSGs as a common endpoint of single-star evolution up to at least $ \sim25\, \rm M_\odot$, associated with type-II SNe due to their hydrogen-rich envelope. While type-II SNe are common \citep[e.g.,][]{Pessi+2025} and RSG SN progenitors have been empirically confirmed \citep[e.g.,][]{Kilpatrick+2023,Xiang+2024}, there is significant evidence that not all RSGs will end their life in this way \citep[e.g.,][]{Das+2025}. 

The majority of massive stars are expected to form in close binary systems, where the two companions will interact at some point during their lifetimes \citep[e.g.,][]{Sana+2012,Sana+2025}, altering the final fate of massive stars \citep[e.g.,][]{Paczynski1967,Vanbeveren+1997}. In addition, the measured extensions of RSG atmospheres have been revealed to be larger than expected \citep[e.g.,][]{Arroyo-Torres+2015} and can range up to $\sim$70\,au \citep{Gonzalez-Tora+2024}, providing ample room for binary interaction even for much larger orbital separations. This was usually not accounted for in previous binary evolution models (where only the photospheric radius is used as an indicator) and thus suggests a paradigm shift in our understanding of binary evolution. 
Indeed, \citet{ercolino24} recently discussed the inclusion of wind Roche-lobe overflow in the binary evolution models and showed that even a very wide companion can play a key role in stripping RSGs of their outermost layers, causing the primary star to spend a significant part of its central He burning with a hotter appearance, possibly performing blue loops in the HR diagram. Recently, several post-RSG supergiants were identified, which evolved to a hotter spectral type following a likely binary interaction \citep[e.g.,][]{Kourniotis+2025, munoz26}, while stripped-envelope and interacting SNe are more common than expected \citep[e.g.,][]{Ercolino+2026}, casting serious doubts on the textbook picture of single RSGs as common massive star evolution endpoints.

Observationally, hot massive companions to RSGs also seem to be rather common. Several have been known since IUE times \citep[e.g.,][]{BussSnow1988,ParsonsAke1998}, and more and more are nowadays identified even in other galaxies within the Local Group \citep[e.g.,][]{Patrick+2025} with hot-companion detection rates ranging up to $40\%$ in M31 and M33 \citep{Neugent+2019, neugent20, neugent21, 2022MNRAS.513.5847P}. Such reported multiplicity, along with the number of detected interacting SNe, marks RSG systems as a crucial intermediate evolutionary stage of massive binary systems, which was largely neglected in the broader astrophysical perspective. Meanwhile, about a third of RSGs also show photometric variations due to long secondary periods \citep[e.g.,][]{kiss06}, which were recently attributed to close-by low-mass companions \citep{goldberg+25}.


\begin{figure*}[t]

\centering 
\includegraphics[width=0.29\textwidth]{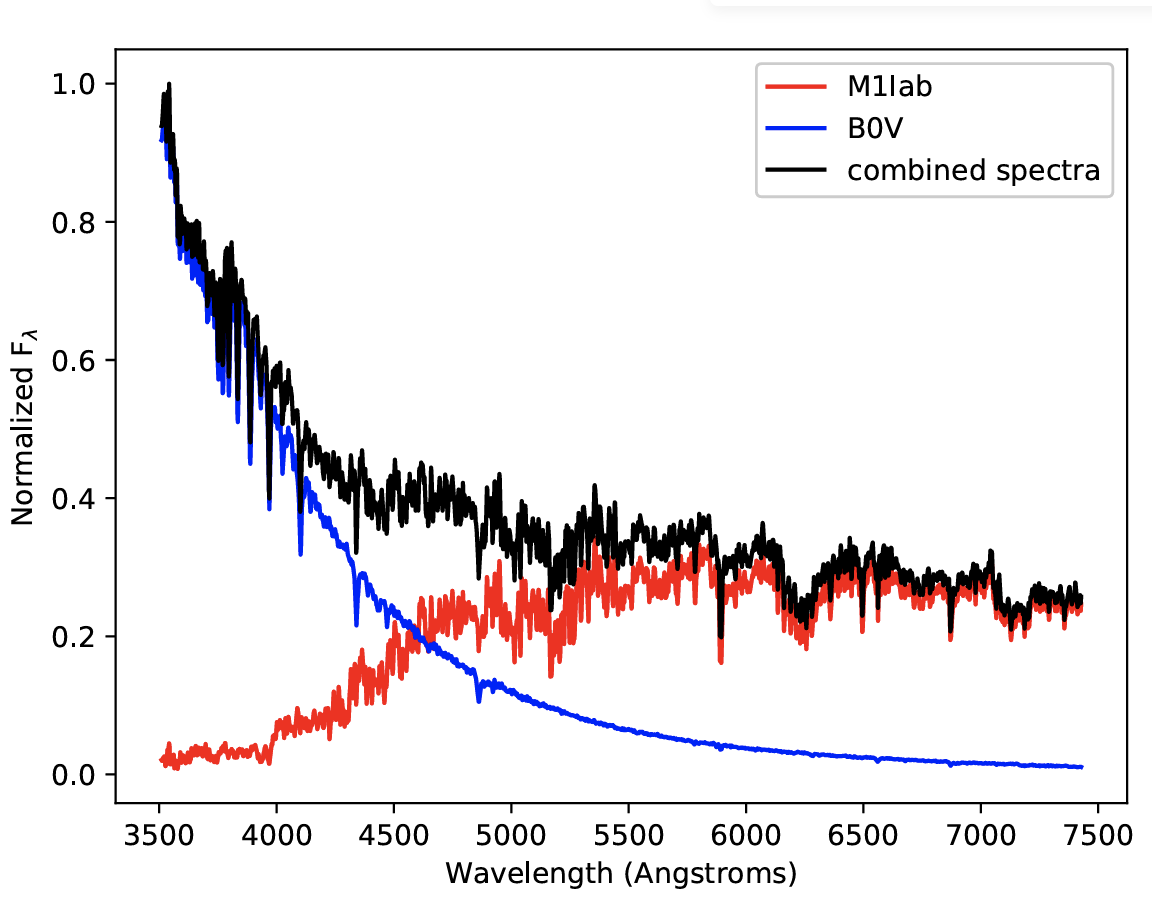} \includegraphics[width=0.7\textwidth]{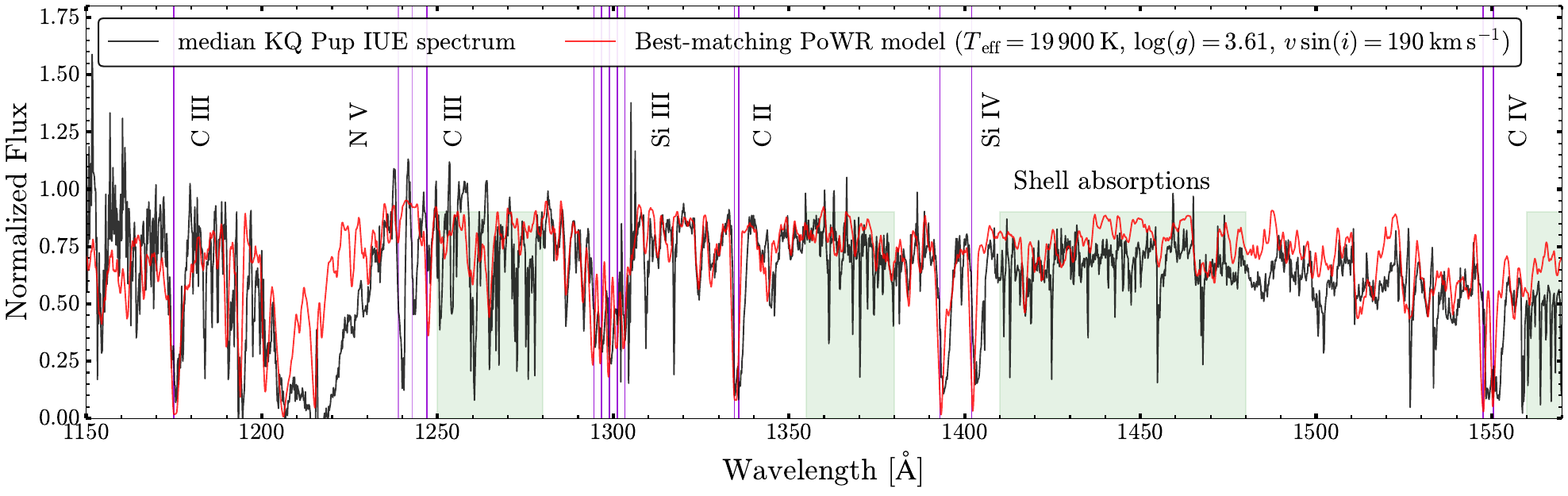} 

\caption{{\em Left panel:} Composite spectrum of an RSG+B binary, demonstrating that at lower wavelengths, the hot companions dominate the spectrum \citep[from][]{Neugent+2019}. {\em Right panel:} Far-UV region of the IUE spectrum of KQ Pup, which shows high-ionization features that cannot form in the RSG wind and are clear B-star signatures \citep[from][]{Jadlovsky+2025}. The spectra also show narrow absorption lines forming in the circumstellar medium of the hot companion and the RSG. \label{fig:specs}} 
\label{fig:fig1}
\end{figure*}

While the frequent existence and importance of hot RSG companions becomes more and more evident, the properties of the hot companion stars and the system configurations remain largely unconstrained \citep{patrick2024}. This is mainly due to the fact that the luminous RSGs dominate the optical light, usually preventing a direct study of the hot companions, while the orbital periods are typically from several years to decades. The few better constrained RSG systems belong to the VV Cephei type binaries \citep{Cowley1969}, where a signature from the hot companion can be seen at longer wavelengths via complex hydrogen emission features which depend on the orbital phase. However, the hydrogen emission lines (e.g., Balmer and Brackett series) arise in the accretion disk of such wind-accreting hot companions \citep{Jadlovsky+2025}, and thus the emission lines cannot be used to characterize properties of the hot companions, nor to determine precise orbital velocity (Fig. \ref{fig:fig2}, left). 

At lower wavelengths ($\lesssim 4000 \rm \:\AA $), the situation is improved as the hot B-type companions begin to dominate the spectral energy distribution \citep[SED, e.g.,][see Fig. \ref{fig:fig1}]{Neugent+2019}. Therefore, high-resolution spectroscopy in the UV with HST is essential, as it is the only observational setup that provides clear spectral diagnostics uncontaminated by the RSG component. 
Nonetheless, only a small number of RSGs were observed with IUE, while this was mostly done in low-resolution ($\sim$7\,\AA) and with low signal-to-noise (S/N). The IUE spectroscopy revealed the companions as B-type stars, but it is insufficient for detailed quantitative spectroscopy and mainly provides information on the SED. Literature reports about the companion properties are thus almost absent, with reported subtypes being rough, flux-based estimates \citep[e.g.,][]{BussSnow1988} rather than verifiable companion information. Traditionally, properties of RSG companions have also been inferred from comparisons with evolutionary models, but this poses the risk of circular reasoning, as it requires an inherent set of assumptions about binary evolution. Notably, even the low-resolution IUE data indicate different luminosity classes for different companions \citep[e.g.,][]{BussSnow1988}, meaning that some B-type companions could be more evolved or have been (partially) stripped. 

Likewise, HST has been used in the past to observe only a few selected RSG systems, such as to study the chromospheres of Betelgeuse \citep{uitenbrook+98, dupree20, dupree+22} 
and Antares \citep{harper26}. For Betelgeuse, it has been hypothesized that a close-by low-mass companion may interact with and modulate its close circumstellar environment \citep{goldberg+25, dupree+26}.  
One of the first binary targets for HST was VV Cephei, the most iconic RSG binary. A dedicated observing campaign was conducted to monitor the system during the eclipse of its hot B component by the RSG \citep{Bauer+2000, Bauer+2007}. This yielded important constraints on the formation regions of the UV spectral features and the colliding winds, as well as the size of the accretion disk. Meanwhile, recent UV surveys in the Local Group discovered hundreds of new RSG+B binaries \citep[e.g.,][]{Neugent+2019,2022MNRAS.513.5847P}. 
Only a handful were characterized with HST \citep{Patrick+2025}. While this work utilized low-resolution HST spectra not sufficient to derive orbital solutions, it demonstrates the power of UV spectroscopy for RSGs, as it yielded important constraints on the system properties, with several systems showing discrepancies between the ages of the RSG and hot components, suggesting past mass transfer, while multiple targets showed signatures of ongoing interaction. The capabilities of HST also allow for taking high-resolution spectra for hot RSG companions in the Magellanic Clouds, as demonstrated by an additional pathfinder programme with the spectral analysis currently underway (L. R. Patrick et al., in preparation). 

So far, only very few RSG systems were monitored with high-resolution HST spectra for longer than a single epoch. While being important to get better insights on the winds and mass-loss processes of RSGs, they thus could not significantly constrain the binary interaction and orbital configurations, nor the detailed properties and evolutionary status of the hot companions, which are both crucial to anchor massive binary evolution models.

\section*{Key science questions which require Hubble's capabilities}

\begin{figure*}[t]
\centering 
\includegraphics[width=0.4\textwidth, keepaspectratio]{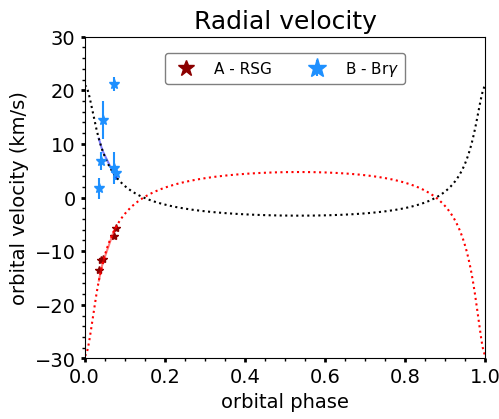} 
\includegraphics[width=0.25\textwidth, keepaspectratio]{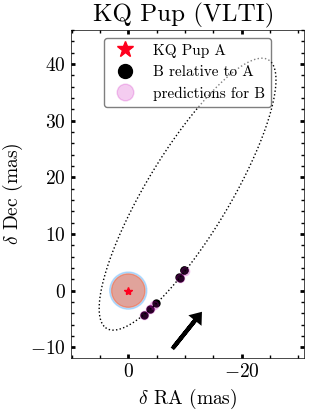} 
\includegraphics[width=0.33\textwidth, keepaspectratio]{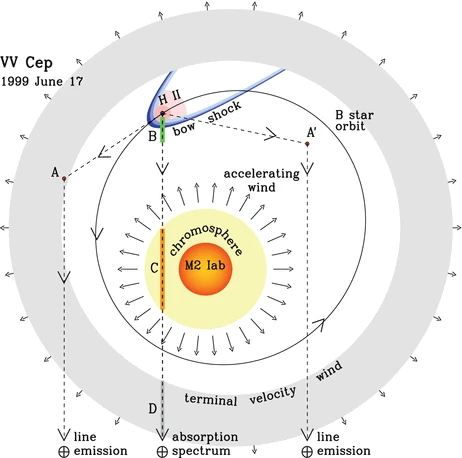}
\caption{
{\em Left and middle panels:} Reconstructed orbit for KQ Pup with VLTI-GRAVITY, based on relative astrometry and radial velocity (RV) with Br$\gamma$ emission line in near-IR \citep{Jadlovsky+2025}. For such systems, a more precise RV from HST using the photospheric lines would add a stronger constraint on the orbital velocity of the B-type companion.
{\em Right panel:} Schematic view of the VV Cep system, showing the orbit and line formation regions \citep{BennetBauer2015}.}
\label{fig:fig2}
\end{figure*}

\noindent%
With its unique UV capabilities, only quantitative HST spectroscopy can deliver accurate stellar and orbital parameters of hot massive companions in RSG binary systems where both components may explode as core-collapse SNe.
Photospheric lines of the hot stars are required to provide complete orbital solutions for the systems, but can only be obtained from UV spectra in RSG+B systems, which require systematic monitoring over years to track orbital periods.
Finally, for interacting systems, the hot companions are embedded in the dense wind of the RSG, 
allowing us to use the B stars as a unique probe of the otherwise hardly directly observable RSG wind properties. 
Therefore, high-resolution HST UV spectroscopy marks the best and only tool to characterise hot companions to RSGs, determine their interaction, and answer major open questions in the evolution of massive stars.
By characterizing stellar and orbital parameters in systems that may ultimately result in double-compact object binary systems, this data is uniquely set to test binary mass-transfer assumptions and constrain wind parameters. 
To address the lack of well-constrained RSG companion properties and binary system configurations, \textbf{multi-year high-spectral-resolution observing campaigns} of a larger sample of RSGs are required. 
No other planned mission in the next decade is foreseen to provide the same capabilities as HST, making the HST the only tool available for such studies. 
Overall, we expect that in the 2030s, HST can answer the following unresolved questions:
\begin{itemize}
  \item \textbf{What are the properties of hot companions to RSGs?} 
  High-resolution UV spectra allow us to determine the stellar and wind properties, plus the chemical composition of hot RSG companions in different metallicity environments, both in our Galaxy and in the Local Group. 
  Meanwhile, model atmosphere codes have developed and can unveil more information about the companions, e.g., FASTWIND \citep{Puls+2005}, CMFGEN \citep{HillierMiller1998}, or PoWR \citep{Graefener+2002}, which can perform a coherent spectral synthesis of photospheric and wind features using non-LTE.

  \item \textbf{How common is multiplicity among companions to RSGs?}  
  Recently, \citet{Jadlovsky+2025} showed that many of the B-type companions may be short-period binaries, suggesting that many RSG systems are hierarchical triples. Only high-resolution HST spectroscopy can confirm such companions and study their properties. 

  \item \textbf{What are the orbital configurations of RSG binary systems?} 
  With the information from the photospheric lines in the high-resolution HST data, the crucial missing piece of information is available to derive an orbital solution for the RSG+B system by combining HST with new data from other missions (see the last Section). The combined orbital solution will also allow us to determine true distances to RSG systems. This knowledge is imperative for a full characterization of such systems, as currently, the available parallax measurements from Gaia are unreliable due to the binary motion, as well as due to the large angular radii of the RSGs.

  \item \textbf{What is the structure of the wind launching region of RSGs?} The presence of a hot star in the vicinity of a RSG provides a unique probe to empirically constrain the inner part of the extended RSG atmosphere \citep[Fig. \ref{fig:fig2}, right, e.g.,][]{HempeReimers1982, Rossi+1992, Patrick+2025}. 
  This will also allow us to better constrain the poorly understood mass-loss process of RSGs \citep{humphreys22}, and the role of companions in driving the asymmetric winds of RSGs \citep{landri24}.

  \item \textbf{What is the role of RSG+B systems in massive star evolution?}
  The B-star properties derived from UV spectroscopy can give unique constraints on the current evolution status of RSG+B systems and 
  -- together with the orbital information -- uncover whether some of the systems match onset configurations of massive common envelope simulations \citep[e.g.][]{Lau+2022} or are progenitors of known Be binaries and high-mass X-ray binaries \cite[e.g.,][]{reig11}. Wind Roche-lobe overflow could also lead to B-type companions becoming Be stars \citep{li26}, or other types of binaries with unclear evolutionary pathways, while interacting RSG systems may evolve towards yellow and blue supergiant systems \citep{munoz26}, connecting to the observed populations of interacting SNe \citep{Ercolino+2026}. From this perspective, interacting RSG systems may represent a rare, short-lived phase of binary interaction, yet they play a crucial role in linking different types of massive binary systems. Lastly, the RSG wind speed information from the high-resolution UV data enables rare constraints on the range of the hydrostatic regime in the extended RSG atmosphere, which is the characteristic quantity for initiating mass transfer in binary evolution models. HST UV spectroscopy of RSG+B systems is therefore a unique probe for both single and binary evolution.

\end{itemize}

\section*{Instrument capabilities and operational requirements}
\noindent%
Higher-resolution spectroscopic modes of STIS and COS in both FUV and NUV (e.g., medium-resolution modes G130M, G160M, G185M, G230M, and high-resolution modes E140H and E230H) are required for disentangling multiple UV line systems and especially for the determination of precise radial velocities (RVs). The FUV spectra show features typical of hot stars (Fig. \ref{fig:fig1}, right), such as broad absorptions of ionized resonance lines (e.g., C\,II, Al\,III) originating in the wind, and also photospheric lines (H and He), allowing us to characterize the companions, whereas the NUV region traces the wind interaction region of the cool and hot component, showing P\,Cygni-type profiles (e.g., Fe\,II). The NUV region is also very useful for constraining the SEDs, as both components contribute comparably to the flux in this region. Lower spectral-resolution modes (e.g., G140L and G230L) are also important for determining SEDs of a larger sample of targets and additional targets that would be too faint for higher-resolution observations.

For extragalactic targets, the higher sensitivity of COS is required, while for the brighter Galactic targets, STIS is sufficient. The small field of view of STIS aperture (25x100 mas) is also exceptionally useful for nearby Galactic RSG+B systems with wide separations, where it will be possible to obtain a separate spectrum of each component, or trace different regions of the wind (e.g., as done for Betelgeuse, \citealt{dupree20}). 

Such high spectral and spatial resolution within the same wavelength range is not foreseen for any other planned missions starting within the next decade, e.g., UVEX (large field of view and low spectral resolution) or the Nancy Grace Roman Space Telescope (optical and IR wavelength region). Therefore, HST will remain our only high-spectral-resolution window into the UV for the 2030s and must remain operational to provide the urgently needed insights on RSG+B systems, such as RV constraints for both long orbital period systems and upcoming newly discovered RSG binary systems from Gaia DR4 and DR5. 


\section*{Towards the development of the HWO}
\noindent%
The Habitable Worlds Observatory (HWO) will be the next flagship mission with pioneering UV facilities. Yet, the orbital periods of RSG binary systems mark a perfect example of why an important missing piece of massive binary evolution cannot and should not wait for HWO to be available. Presently, the orbital period distribution is not known for the large majority of RSG+B systems, although the interacting RSG systems have orbital periods within the range of about 3 to 30 years \citep{patrick2024}. With an extension of HST's lifetime throughout the 2030s, RV measurements of hot RSG companions would allow us to determine the full orbits of systems with periods between 3 and 10 years, and constrain a significant part of the orbit for systems with longer periods. This would be sufficient to determine the dynamical masses and mass ratio in many cases, while systems with longer orbital periods could be constrained by additional observations with HWO observations. 
The forthcoming Gaia data releases will further detect many more systems. For these systems, the availability of HST in the 2030s is crucial to provide an important baseline to later follow up on such systems with HWO. 


While HST will be able to constrain the properties and orbits of hot companions to RSGs in the Milky Way and the Magellanic Clouds, only the HWO will enable us to study these systems in further away galaxies and with a wider range of metallicities. Yet, HWO will not be available for at least another two decades, while the specifics of their instruments have to be defined now. 
The proposed HST legacy observations on RSG binary candidates will generate the sample necessary to define the precise requirements that HWO needs to meet in order to successfully probe the UV-bright companions in RSG systems in other galaxies.


\section*{Hubble Legacy Programs enabled by synergies with other instruments:}
\noindent%
In the starting era of high-angular resolution interferometry and high-precision astrometry, the unique access to the UV provided only by HST enables a pioneering opportunity to uncover the physics of massive binary evolution imprinted in RSG binary systems. Synergies to be exploited in particular arise from

\begin{itemize}
\item \textbf{Optical interferometry (VLTI/CHARA):} Recent updates to high-angular resolution interferometric instruments, such as VLTI/GRAVITY+, provide better sensitivity and 
improved astrometric precision down to $\sim 10 \mu \rm as$. Along with the new high-spectral-resolution and high-contrast capabilities, we can detect fainter companions than before, study their interaction, and trace their orbits \citep[e.g., Fig. \ref{fig:fig2}, middle,][]{Jadlovsky+2025}. 
Combining these observations with quantitative HST UV spectroscopy allows the orbital configurations to be fully solved.
Such a combination would yield dynamical masses for RSGs and their companions to a level of precision that is currently unattainable through other observational techniques. 

\item \textbf{Gaia astrometry:}
The Gaia data release 4 (DR4; expected December 2026) and the later DR5 will provide multi-epoch astrometric and RV measurements over a 5-10 year baseline for thousands of RSGs in our Galaxy and the Magellanic Clouds and will result in hundreds of newly discovered binary systems. 
HST will play a key role in characterizing newly discovered hot companions in the next 10 years and is the only facility that can deliver the quantitative stellar and orbital parameters required to understand the evolution and final fate of these systems. Besides constraining the orbits of luminous companions, deep HST observations will be necessary to confirm or rule out candidates for compact companions identified by Gaia.

\end{itemize}

To capitalize on these synergies, the availability of HST throughout the 2030s is essential, as it will remain our only way to study photospheric lines of the hot companions in the UV for at least another decade. 
We therefore propose a legacy program taking high-resolution FUV and NUV spectra of a statistically meaningful sample of interacting RSGs to constrain the orbital periods and mass ratio distribution of RSG binary systems. For this, long-term monitoring of at least $60$ binary RSGs -- $\gtrsim 20$ in the Milky Way and each of the Magellanic Clouds -- is required, probing different metallicities. As the typical periods span from several years to decades, multi-epoch observations over the whole 2030s are crucial for proper orbital constraints and will provide an unprecedented legacy dataset for understanding RSGs, their winds, and massive binary evolution. Accurate masses and orbital configurations will have a significant impact on our ability to connect the observed RSG populations in the local Universe to core-collapse SNe and their progenitors.




\newpage
\bibliography{sample701}{}
\bibliographystyle{aasjournalv7}



\end{document}